\documentclass{jpp}
\usepackage{graphicx}
\usepackage{physics}
\usepackage[utf8]{inputenc}
\usepackage[T1]{fontenc}
\usepackage{amsmath}
\usepackage{float}
\usepackage{hyperref}
\usepackage[dvipsnames]{xcolor}
\usepackage{siunitx}

\shorttitle{Runaway Dynamics in Reactor-Scale Spherical Tokamak Disruptions}
\shortauthor{E. Berger and others}

\title{Runaway Dynamics in Reactor-Scale Spherical Tokamak Disruptions}

\author{Esm\'{e}e Berger\aff{1}, 
Istv\'{a}n~Pusztai\aff{1}
  \corresp{\email{pusztai@chalmers.se}},
  Sarah L. Newton\aff{2},
    Mathias Hoppe\aff{3},
  Oskar Vallhagen\aff{1},
  Alexandre Fil\aff{2},
 \and   T\"{u}nde F\"{u}l\"{o}p\aff{1}}

\affiliation{\aff{1}Department of Physics, Chalmers University of Technology, G\"{o}teborg, SE-41296, Sweden
\aff{2}  Culham Centre for Fusion Energy, Abingdon, Oxon OX14 3DB, United
Kingdom
\aff{3} Swiss Plasma Center, Ecole Polytechnique F\'{e}d\'{e}rale de Lausanne, Lausanne, CH-1015, Switzerland
}

\begin{document}

\maketitle

\begin{abstract}
Understanding generation and mitigation of runaway electrons in disruptions is important for the safe operation of future tokamaks. In this paper we investigate runaway dynamics in reactor-scale spherical tokamaks. We study both the severity of runaway generation during unmitigated disruptions, as well as the effect that typical mitigation schemes based on massive material injection have on runaway production. The study is conducted using the numerical framework \textsc{dream} (\textit{Disruption Runaway Electron Analysis Model}). We find that, in many cases, mitigation strategies are necessary to prevent the runaway current from reaching multi-megaampere levels. Our results indicate that with a suitably chosen deuterium-neon mixture for mitigation, it is possible to achieve a tolerable runaway current and ohmic current evolution. With such parameters, however, the majority of the thermal energy loss happens through radial transport rather than radiation, which poses a risk of unacceptable localised heat loads.

\end{abstract}

\section{Introduction}
\label{sec:intro}

 \textit{Spherical tokamaks} (STs) have a significantly smaller aspect ratio, i.e.~ratio of plasma major radius ($R_0$) to plasma minor radius ($a$), than conventional tokamaks. Their compact shape allows for more efficient confinement at a given magnetic field strength than in conventional tokamaks. The more compact configuration can lower the construction cost and spherical tokamaks have been proposed as component testing facilities, to aid the development of magnetic confinement fusion. However, beyond their use in component testing, there is an effort to construct STs suitable for energy production in order to accelerate the path to commercially available fusion power. Part of this effort is the \textit{Spherical Tokamak for Energy Production} (STEP) program in the UK, aiming to design and construct a prototype fusion energy plant by 2040 \citep{stepbook,STEPaim}. One phase in the STEP program has been to develop a preliminary high power ST design to understand the interplay between turbulence and shaping for a STEP reactor equilibrium. This design is called BurST, short for \textit{Burning Spherical Tokamak} \citep{Bhavin_phd}.

One of the remaining challenges of reactor-scale tokamaks is the rare, yet potentially detrimental, occurrence of rapid, unwanted degradation of the plasma magnetic confinement, and associated loss of thermal energy, known as a \textit{disruption}. In the first phase of a disruption there is a dramatic decrease of the plasma temperature from the initial $\sim\SI{10}{\kilo\electronvolt}$ down to $\sim \SI{10}{\electronvolt}$, within a few milliseconds. This phase is called the \textit{thermal quench} (TQ), and it is caused by a combination of an elevated radial transport, due to instabilities temporarily destroying magnetic flux surfaces, and atomic physics processes, including radiation by impurities and plasma dilution. Impurities may enter the plasma unintentionally, or could be injected as part of a disruption mitigation scheme. 
 Even though a part of the plasma thermal energy is isotropically radiated away, the remaining fraction, which is transported to the tokamak wall, can be significant. These transported heat loads tend to be very localised, and may damage the plasma-facing components. 

As the temperature in the plasma drops during a disruption, the plasma resistivity increases,  leading to a decay of the plasma current and characterising the beginning of the second phase of a disruption --- the \textit{current quench} (CQ). As a consequence, currents are induced in the reactor wall leading to structural forces that can be large enough to damage the device. Furthermore, an electric field is induced in the plasma when the plasma current drops, which, if strong enough, can accelerate electrons to relativistic energies. Such electrons are called \textit{runaway electrons} and they can severely damage areas upon which they have an uncontrolled impact.

When a disruption occurs, its impact will have to be mitigated so that the tokamak does not suffer substantial damage. The runaway current should be below a certain limit  in order to avoid unacceptable localised melting of the wall, and possibly also underlying structures; here we require this limit to be comparable to that in ITER, \SI{150}{\kilo\ampere} \citep{Lehnen_talk} {(note that this value is approximate, as the damage caused depends on the localization of the RE beam impact)}. The current quench time 
$t_\text{CQ}$, i.e.~the time it takes for the ohmic component of the current to decay,
should also be in an acceptable range to avoid {excessive mechanical stresses due to} eddy currents and halo currents in the wall. The lower limit is expected to be around \SI{20}{\milli\second}, with a preliminary upper limit approximately \SI{100}{\milli\second} \citep{Hender_private_comm}{. For reference, the upper limit on ITER is slightly more relaxed; \SI{150}{\milli\second} \citep{Hollmann}}. Furthermore, the fraction of the thermal energy lost through radial transport {during the TQ} should remain below {a certain value, to avoid unacceptable localised heat loads on the wall. The ITER target for the upper limit of the transported fraction, \SI{10}{\percent} \citep{Hollmann} is also applied here.}

One of the methods proposed to mitigate these potentially harmful effects is massive material injection \citep{Hollmann}. Material injection can act to reduce the runaway generation, as the critical energy for electron runaway is higher at an elevated electron density. A suitable material to inject for this purpose is deuterium. Massive material injection can also be used to control the current quench time, as $t_\text{CQ}$ is proportional to the conductivity which depends on the final temperature after the TQ, and this temperature is in turn determined by an equilibrium between the ohmic heating and impurity radiation. A suitable material to inject for this purpose is thus a radiating impurity species, typically noble gases such as neon or argon. The injected material can also radiate away a large fraction of the thermal energy during the disruption. To accommodate all requirements at once, injection of a mixture of deuterium and a noble gas is preferred. This has the added benefit that the radiation efficiency is enhanced through the increase in electron density offered by the injected deuterium, as the collisional excitation rate depends on the electron density.

Runaway electron generation has been studied extensively for conventional tokamaks. To date, runaways have rarely been observed in spherical tokamaks (STs) during the short disruption timescales in the current small devices. The disruption dynamics usually differ from those in a conventional tokamak \citep{Gerhardt_2009,Thornton_phd} and it is therefore not straightforward to transfer the results about runaway electron dynamics in conventional tokamaks to STs. ST plasmas are typically strongly elongated, and it has been shown that elongated plasmas in a conventional tokamak produce fewer runaway electrons during disruptions \citep{elongation}. The aim of this paper is therefore to numerically investigate the potential runaway dynamics in reactor-scale spherical tokamaks, using input parameters for the BurST reactor design as a basis.

\section{Disruption and Runaway Modelling}
 \label{sec:methods}

 The results are obtained using the numerical framework \textsc{dream} (\textit{Disruption Runaway Electron Analysis Model}) \citep{DREAM} that self-consistently evolves background plasma parameters together with the runaway dynamics during a disruption. For our purposes, \textsc{dream} is used in fluid mode, in which the thermal electron bulk, the runaway electrons and the ion species are each treated as fluid species. The various physics mechanisms activated in fluid mode, such as the runaway generation rates, show good correspondence to the more sophisticated kinetic results, with the model included for processes such as Dreicer generation constructed from large  kinetic simulation databases \citep{NN_Dreicer}. We do not require inherently kinetic outputs, such as the phase space distribution function, in this scoping study of ST-based reactor scale disruptions and can thus take advantage of the much less computationally demanding fluid mode to perform wide parameter explorations. 

In the fluid model, the thermal electron bulk is characterised by its density $n_\text{e}$, temperature $T_\text{e}$, and the ohmic current density parallel to the magnetic field lines,  $j_\text{ohm}$. In our case, $n_\text{e}$ represents the density of all free electrons that are not runaways, i.e.~$n_\text{e} = n_\text{free} - n_\text{re}$, and as such it is determined by the evolution of the runaway and free electron densities. The free electron density is determined by the ion composition of the plasma. The density $n_i^{(j)}$ of ion species $i$ with charge state $j$ is evolved according to the ion rate equation
\begin{equation}
    \pdv{n_i^{(j)}}{t} = n_\text{e}[\mathcal{I}_i^{(j-1)}n_i^{(j-1)} + \mathcal{R}_i^{(j+1)}n_i^{(j+1)} - (\mathcal{I}_i^{(j)}+\mathcal{R}_i^{(j)})n_i^{(j)}],
\end{equation}
where $\mathcal{I}$ and $\mathcal{R}$ are ionisation and recombination rates, respectively, which depend on the plasma parameters \citep{iter_matinj}. The ions are assumed to be fully ionised at the start of the simulation, as the plasma is assumed to be representative of steady-state operation before the disruption occurs.  

The time evolution of the runaway electron density $n_\text{re}$ is given by
\begin{equation} \label{eq:re_growth}
    \pdv{n_{\text{re}}}{t} =  \gamma_{\text{Dreicer}} + \gamma_{\text{hot-tail}} + \gamma_{\text{tritium}} +  \Gamma_{\text{ava}}n_{\text{re}}+\frac{1}{V'}\frac{\partial}{\partial r}\left[V' D_{\rm re} \frac{\partial n_{\text{re}}}{\partial r} \right].
\end{equation}
Here, each source term marks the generation rate of the mechanism indicated by its subscript.  Dreicer runaway generation is a phenomenon where electrons diffusively leak into the runaway region due to small-angle collisions \citep{Dreicer1959}. The hot-tail generation mechanism produces runaways due to the fastest electrons not having time to thermalise before the electric field rises, after a sufficiently fast temperature drop during the TQ \citep{Helander2004,Smith2005,Svenningsson_PRL}. Tritium in the device undergoes $\beta$-decay, generating energetic electrons according to a continuous energy spectrum, a part of which may be in the runaway region \citep{MartinSolis2017,elongation}. Runaway electrons generated by Compton scattering of $\gamma$-rays from the activated wall  are not included here, due to the lack of input data for the spectrum of $\gamma$-photons emitted from the plasma-facing components in a reactor such as BurST. Finally, close collisions between a runaway electron and a thermal one can transfer sufficient energy to the latter so that it also becomes a runaway electron. This leads to an exponential increase in the number of runaway electrons, with the avalanche multiplication rate $\Gamma_{\text{ava}}$ \citep{Rosenbluth_Putvinski,Embreus2018}. The last term of Eq.~(\ref{eq:re_growth}) describes a diffusive radial transport of REs. For the transport of REs and heat (to be discussed in more detail later), we use a collisionless Rechester-Rosenbluth-type diffusion coefficient \citep{Rosenbluth_D} of the form 
\begin{equation}
    D = \pi \left| v_\| \right| R_0 \left(\delta B/B\right)^2,
    \label{generalDiff}
\end{equation} 
for particles with parallel velocity $v_\|$, with the normalized magnetic perturbation amplitude $\delta B/B$. When evaluating $D_{\rm re}$, we assume $v_\|=c$ for all REs. The RE particle transport term is activated only in some simulations with material injection.

Existing analytical expressions for the Dreicer runaway generation rate $\gamma_{\text{Dreicer}}$  neglect effects of partial screening, which have been shown to be important \citep{Hesslow_CoulLog}. We therefore employ the neural network trained on a large number of kinetic simulations presented in \citep{NN_Dreicer}, which take effects of partial screening into account.
The energy dependent model for the Coulomb logarithm is described in equation~(18) in \citep{DREAM}, which in the fluid mode is evaluated at a representative runaway momentum of $20\,m_\text{e}c$.
The model for the tritium decay generation is taken to be as in \citep{elongation}.
In the model for the hot-tail, an analytic approximate distribution function and critical runaway momentum are calculated as functions of the background plasma parameters and electric field. These are then used to evaluate $\gamma_{\text{hot-tail}}$, as described in Appendix C.4 of \citep{DREAM}.  The avalanche multiplication rate is described in \citep{Hesslow_fluid_ava}, and it accounts for both partial screening and magnetic trapping effects. 
Magnetic trapping is likely substantial at the outer flux surfaces in tight aspect ratio devices if the collisionality is low. This is accounted for by multiplying the avalanche growth rate with the effective passing fraction defined in equation (B.16) of \citep{DREAM}
in the simulations of unmitigated disruptions. 

The magnetic geometry is parameterised according to the analytical model described by \citet{Miller98}. In this geometry model, the radial coordinate, $r$, measures the half width of a flux surface in the mid-plane. The flux surfaces are parameterised by their elongation $\kappa(r)$, Shafranov shift $\Delta(r)$, triangularity $\delta(r)$, 
and toroidal magnetic field function $G(r)=R B_\varphi(r)$. There is however one difference compared to the original Miller model, namely that the Shafranov shift is defined to be zero at the magnetic axis \citep{DREAM}.  Apart from specifying the above geometrical parameters, the user also inputs the plasma minor radius $a$, major radius (at the magnetic axis) $R_0$, and wall radius $b = a + r_\text{wall}$, where $r_\text{wall}$ is the distance between the plasma edge and the wall\footnote{The wall radius is representative of the location of the toroidally closed conducting structure closest to the plasma in terms of poloidal magnetic flux, and as such it is not necessarily the distance to the first wall.}. 

The evolution of the electron temperature $T_\text{e}$ is prescribed as an exponential decay in the simulations of unmitigated disruptions, whereas the temperature is self-consistently evolved for the mitigated disruptions. The exponential temperature decay is given by 
\begin{equation}\label{eq:expT}
    T_{\text e}(t,r) = T_f(r) + [T_0(r) - T_f(r)]\text{e}^{-t/t_0}, 
\end{equation}
where $T_0(r)$ is the initial temperature profile, $t_0$ is the decay time scale and $T_f(r)$ is the final temperature profile.  After a disruption, the final temperature $T_f$ is usually flatter than the initial $T_0$-profile and is therefore taken as a radially constant value, for simplicity.
The self-consistent temperature evolution, used in the mitigation simulations, is described by the energy balance equation for the thermal energy density of the bulk electrons $W_\text{e}$, which relates to the temperature through $W_\text{e} = 3n_\text{e}T_\text{e}/2$. The energy balance equation takes into account ohmic heating by the electric field, electron heat diffusion, bremsstrahlung radiation losses, line radiation losses, and ionisation energy losses \citep{DREAM}.
In the mitigation simulations, the injected material is assumed to be present as a neutral gas at the beginning of the disruption, with a radially constant profile. 

In the energy balance equation governing temperature evolution, heat transport is included during the material injection simulations. The heat diffusion coefficient is obtained by taking the heat flux moment of (\ref{generalDiff}) for a Maxwellian distribution, yielding  $ D_W \approx 2\sqrt{\pi} v_\text{th,e} R_0 \left(\delta B/B\right)^2$. Whenever heat and RE particle transport are both active, both diffusivities are calculated using the same magnetic perturbation amplitude, $\delta B/B$, for consistency.
Furthermore, the effect of opacity is included in the mitigation simulations by using ionisation, recombination and radiation rates for the hydrogen isotopes that are based on the assumption of the plasma being opaque to Lyman radiation. This has been shown to significantly affect the results by reducing excessive cooling and recombination, and thereby reducing the avalanche growth rate \citep{Vallhagen_SPI}. 

The total current density is given by $j = j_\text{ohm} + j_\text{re}$, where $j_ {\text{re}} =e c n_{\text{re}}$ as the runaway electrons are assumed to move with the speed of light $c$ parallel to the magnetic field \citep{DREAM}. The evolution of $j$ is governed by the evolution of the poloidal flux $\psi(r)$ \citep{DREAM,Pusztai_curr_rel}. In this evolution, the electrical conductivity $\sigma$ enters, for which we employ the model  described by \citet{Redl}, that takes into account the effects of trapping, and is valid for arbitrary plasma shaping and collisionality. The boundary conditions for the evolution equation of $\psi(r)$ take into account currents in the passive structures surrounding the plasma, denoted by $I_\text{wall}$, via the following set of equations \citep{Pusztai_curr_rel}
\begin{align}
    \psi(a) &= \psi(b) - MI_\text{tot},\\
    \psi(b) &= -L_\text{ext}[I_\text{tot} + I_\text{wall}], \\
    V_\text{wall} &= R_\text{wall}I_\text{wall}.
\end{align} 
Here,  {$I_\text{tot}$ is the total plasma current,}  $M$ is the plasma-wall mutual inductance, $L_\text{ext}$ is the external inductance, $R_\text{wall}$ the wall resistance, and $V_\text{wall}$ the loop voltage in the conducting structure. $M$ is calculated internally in \textsc{dream}, whereas $L_\text{ext}$ and $R_\text{wall}$ are determined by the user and used to specify a resistive timescale of the wall, $t_\text{wall}=L_\text{ext}/R_\text{wall}$. The user thus determines the wall response model by specifying $t_\text{wall}$, as well as the wall radius $b$.

When studying the evolution of the current components during the CQ, we consider specifically the fraction of the initial current converted to runaways, as well as the decay time scale of the ohmic current.  The current conversion is defined as the maximum runaway current during the simulation divided by the initial plasma current $
    \text{CC}=\max(I_{\text{re}})/I_{\text{tot},0}$.
The reason for taking the maximum runaway current as opposed to the runaway current at the end of the simulation, which is not necessarily the maximum, is that runaways can be lost to the wall at any time during a disruption. This means that with this definition of current conversion we obtain the worst case scenario for each simulation. The current quench time will be  calculated as 
$ t_\text{CQ} = \left[ t(I_\text{ohm} = 0.2I_\text{tot,0}) - t(I_\text{ohm} = 0.8I_\text{tot,0})\right]/0.6$ \citep{Gerhardt_2009}.
In all simulations, unless stated otherwise, the simulation is ended  \SI{150}{\milli\second} after the beginning of the disruption. {This value is also inspired by previous ITER simulations, e.g.~\citep{iter_matinj}, being comparable with the timescale of the disrupted plasma vertically drifting into the wall in ITER \citep{Hollmann}.}

Using the model detailed here, we demonstrate trends and dependencies of the runaway behaviour in reactor-scale STs in the following sections.
Quantitative predictions will require some of the above modelling assumptions to be lifted, for example, suitable inclusion of the Compton scattering source, but the runaway levels found in unmitigated disruptions and the responses to material injection already indicate directions which should be pursued.

\section{Unmitigated Runaway Dynamics}
\label{sec:results}
In this section we study the severity of runaway generation in unmitigated disruptions in reactor-scale STs.
We first turn our attention to the effect of the temperature decay during the TQ on the evolution of the current components in the subsequent CQ, to explore the scale of the runaway problem depending on the decay time scale $t_0$ and the final temperature $T_f$. We then illustrate the underlying runaway dynamics for a baseline case.  Finally, the sensitivity of the results for the baseline case to a number of parameter and model choices is investigated. 

{The simulations assume a plasma with major and minor radius of $R_0=3.05\,\rm m$ and $a=1.5\,\rm m$, respectively, with a core electron temperature of  $20 \,\rm keV$, and density of $10^{20}\,\rm m^{-3}$, and a total plasma current of $21\,\rm MA$. The elongation of the outermost flux surface is $\kappa(a)=2.8$. More details on the  BurST plasma and magnetic geometry profiles are provided in the Appendix.}

\subsection{Temperature Decay Parameter Scan}
\label{sec:res_tempscan}
\begin{figure}
    \begin{center}
    \includegraphics[width=\textwidth]{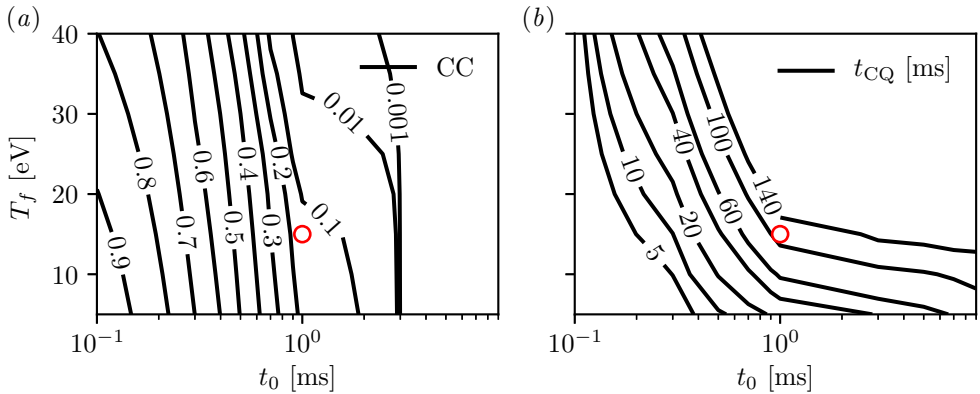}
    \end{center}
    \vspace{-\baselineskip}
    \caption{Contours of (\textit{a}) current conversion (CC) and (\textit{b}) CQ time $t_{\text{CQ}}$ as functions of decay time scale $t_0$ and final temperature $T_f$ of the exponential temperature decay in (\ref{eq:expT}). The red circle marks the case where $t_0=\SI{1}{\milli\second}$ and $T_f=\SI{15}{\electronvolt}$, our baseline case, which is studied in more detail in \ref{sec:res_baseline}. Apart from $t_0$ and $T_f$ the parameters are the same as in the baseline case.}
    \label{fig:TQscan}
\end{figure}

The temperature evolution is modelled by an exponential decay as described by equation~(\ref{eq:expT}). We scan over the experimentally expected ranges $t_0=\SI{0.1}{\milli\second}-\SI{10}{\milli\second}$ and $T_f=\SI{5}{\electronvolt}-\SI{40}{\electronvolt}$. Note that $t_0$ and $T_f$ are not something that one can ``choose'', as they depend on the transport and atomic physics at play, so the results in figure~\ref{fig:TQscan} indicate the severity of runaway generation in reactor-scale STs depending on the temperature decay parameter values. 

In figure~\ref{fig:TQscan}\textit{a} we see that the runaway production is high when the cooling is rapid ($t_0<\SI{1}{\milli\second}$), while the dependence on the temperature reached after the TQ is not very strong in this region. For slower cooling rates ($t_0>\SI{1}{\milli\second}$), however, the current conversion can differ by an order of magnitude depending on the final temperature, with higher values for lower temperatures. The region where the runaway generation is the least problematic for unmitigated disruptions is at $t_0>\SI{3}{\milli\second}$, where the runaway current would be lower than \SI{21}{\kilo\ampere}, i.e.~well below the \SI{150}{\kilo\ampere} limit. In the rest of the $t_0 - T_f$ space some form of mitigation would be necessary to keep the maximum runaway current below this limit. 

Comparing figure~\ref{fig:TQscan}\textit{b} to figure~\ref{fig:TQscan}\textit{a}, we see that the current conversion is in general high when $t_{\text{CQ}}$ is short. Furthermore, we see in figure~\ref{fig:TQscan}\textit{b} that in the region where the current conversion is the lowest ($t_0>\SI{3}{\milli\second}$), the CQ time is strongly dependent on the final temperature, such that $t_{\text{CQ}}$ is below the \SI{100}{\milli\second} or \SI{150}{\milli\second} limit only if the plasma temperature falls below approximately \SI{15}{\electronvolt} after the disruption. This means that in order to satisfy both the demands on the current conversion as well as on the CQ time, mitigation strategies would be necessary not only when $t_0<\SI{3}{\milli\second}$, but also when $T_f\gtrsim\SI{15}{\electronvolt}$ in the region where $t_0>\SI{3}{\milli\second}$.

\subsection{Baseline Case}
\label{sec:res_baseline}

\begin{figure}
    \begin{center}
    \includegraphics[width=\textwidth]{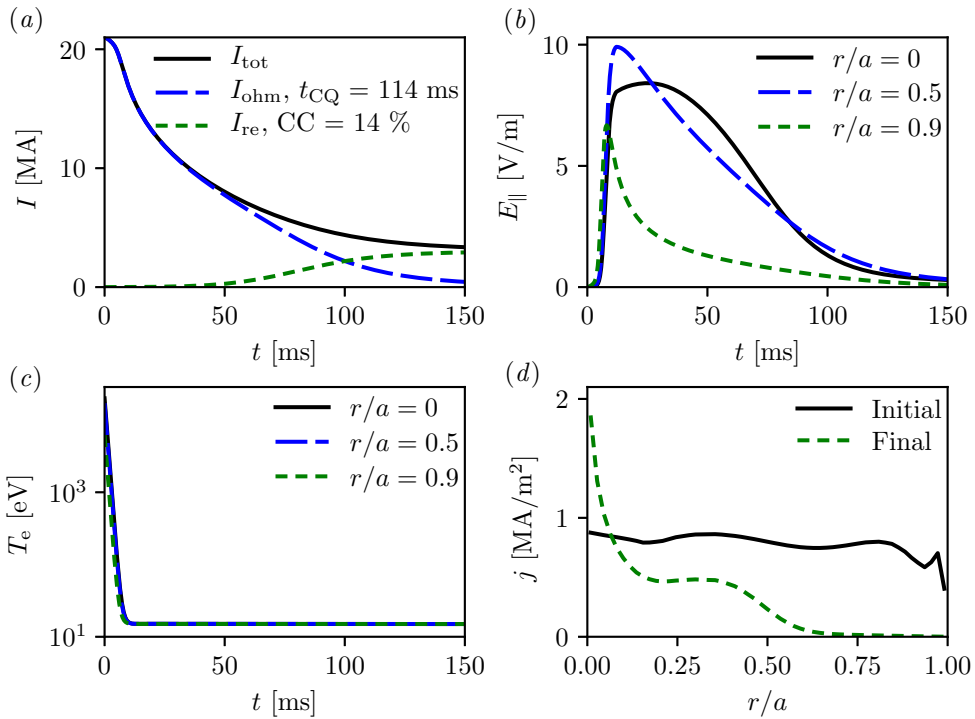}
    \end{center}
    \vspace{-\baselineskip}
    \caption{Plasma current, electric field and temperature evolution in the baseline case where $t_0=\SI{1}{\milli\second}$ and $T_f=\SI{15}{\electronvolt}$. (\textit{a}) Total plasma current (solid) as function of time, together with the ohmic (long dashed) and runaway (short dashed) contributions. (\textit{b}, \textit{c}) Electric field and electron temperature evolution at different radii, given in the legend. (\textit{d}) Initial and final radial current density profiles.} 
    \label{fig:1-15}    
\end{figure}

\begin{figure}
    \begin{center}
    \includegraphics[width=\textwidth]{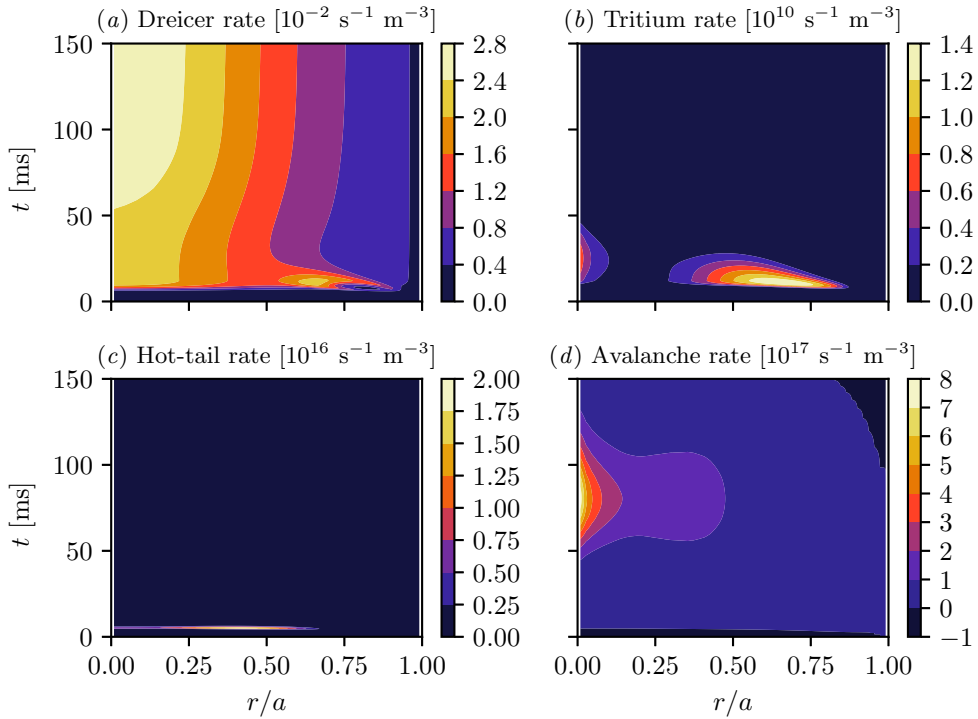}
    \end{center}
    \vspace{-\baselineskip}
    \caption{Time evolution of the radial distributions of the (\textit{a}) Dreicer, (\textit{b}) tritium decay, (\textit{c}) hot-tail and (\textit{d}) avalanche runaway rates, in the baseline case where $t_0=\SI{1}{\milli\second}$ and $T_f=\SI{15}{\electronvolt}$. Note the different scales indicated in the sub-figure headings.} 
    \label{fig:1-15-mech}
\end{figure}
In order to gain more insight into the runaway dynamics at play, we take a close look at a case near the centre of the scanned parameter space, marked with a red circle in figure~\ref{fig:TQscan}. The parameters of this baseline case are: $t_0=\SI{1}{\milli\second}$, $T_f=\SI{15}{\electronvolt}$, 50\% tritium, $r_{wall}=0$ and $t_{wall}=\infty$ (perfectly conducting wall); the plasma current, electric field and temperature evolution are shown in figure~\ref{fig:1-15}. The time evolution of the radial distributions of runaway rates for the different mechanisms are plotted in figure~\ref{fig:1-15-mech}.
  
The current conversion at these temperature decay parameters is \SI{14}{\percent}, meaning that this disruption would not be tolerable without mitigation due to the large runaway current generated. Figure~\ref{fig:1-15}\textit{b} shows that the maximum of the electric field is about \SI{10}{\volt\per\meter}, which is not sufficient to make the Dreicer contribution to the runaway generation significant, as is evident {from figure~\ref{fig:1-15-mech}\textit{a}.} The runaway production is the largest on axis but also has contributions for larger radii, as shown in figure~\ref{fig:1-15}\textit{d}. This differs from similar results for tokamaks with a more conventional shape, where the production only has the on-axis peak and approximately exponentially drops to zero for larger radii \citep{elongation}. 

In {figure~\ref{fig:1-15-mech}\textit{c}} we see that the dominant primary generation mechanism is hot-tail, as its maximum value is about six orders of magnitude larger than that of the second most important mechanism, the tritium decay shown in {figure~\ref{fig:1-15-mech}\textit{b}}. As the hot-tail generation happens very early during the disruption, where transport may be strong due to a high level of magnetic fluctuations, this result is likely to overestimate the hot-tail seed at the end of the TQ because transport has not been taken into account. When completely disabling the hot-tail seed, but otherwise using identical settings, the runaway conversion plummets to \SI{0.0001}{\percent}, see table \ref{Tab:cc-tcq}, thus \SI{14}{\percent} can be taken as an upper limit for the current conversion in this case. Note, however, that $t_{\text{CQ}}$ is longer in the case where hot-tail generation is excluded and the current quench is incomplete after \SI{150}{\milli\second}, with about \SI{2.1}{\mega\ampere} ohmic current remaining. The residual ohmic current could potentially convert to runaway current beyond this point. However, when running this simulation for twice as long, i.e.~until \SI{300}{\milli\second}, the resulting current conversion is instead \SI{0.0005}{\percent}. This means that the runaway current is about \SI{0.1}{\kilo\ampere}, still well below the limit, and the remaining ohmic current is about \SI{0.3}{\mega\ampere} --- so if the entire hot-tail seed would be lost through radial transport this is a disruption where mitigation would not be necessary.

The overall dominant runaway mechanism is avalanche multiplication, with a maximum value of about two orders of magnitude larger than the hot-tail maximum, and with high rates for a much longer period of time, as seen in {figure~\ref{fig:1-15-mech}\textit{d}}. As the avalanche gain increases exponentially with the initial plasma current, it is no surprise that avalanche is the dominant runaway generation mechanism in this case, where there is a high runaway seed and initial plasma current. Due to the dominance of the avalanche, the total runaway generation rate is almost identical to the avalanche rate and is therefore not shown separately. Also, it can be noted that there is a region with negative avalanche multiplication, which occurs when the electric field goes below the critical electric field \citep{Hesslow2018}. A runaway electron can then lose enough energy when colliding with the thermal electron bulk that it falls out of the runaway region, without knocking the thermal electron into the runaway region, and so the runaway density decreases in time. 

 \subsection{Sensitivity Study}
\label{sec:res_sensitivity}
In table~\ref{Tab:cc-tcq} we summarise the current conversion factor, the current quench time and the remaining ohmic current for a number of different cases, including the baseline (entry at the top). As expected, the influence of tritium decay on the runaway rate is weak as evident from table~\ref{Tab:cc-tcq}; neither the current conversion nor the CQ time is affected when changing the initial tritium content in the plasma between the two limiting cases of pure deuterium and pure tritium.
\begin{table}

    \begin{center}
   \def~{\hphantom{0}}
        \begin{tabular}{lccc}
            \hline
             Case & Current conversion & $t_{\text{CQ}}$ & Remaining $I_{\text{ohm}}$ \\[3pt]     
             Baseline & \SI{14}{\percent} & \SI{114}{\milli\second} & \SI{0.44}{\mega\ampere} \\ 
             No hot-tail & \SI{0.0001}{\percent} & \SI{145}{\milli\second} & \SI{2.1}{\mega\ampere} \\
             \SI{0}{\percent} tritium & \SI{14}{\percent} & \SI{114}{\milli\second} & \SI{0.44}{\mega\ampere} \\ 
             \SI{100}{\percent} tritium & \SI{14}{\percent} & \SI{114}{\milli\second} & \SI{0.44}{\mega\ampere} \\ 
             No shaping & \SI{30}{\percent} & \SI{41}{\milli\second} & \SI{0.07}{\mega\ampere} \\
             No trapping & \SI{24}{\percent} & \SI{86}{\milli\second} & \SI{0.22}{\mega\ampere} \\
             $r_{\text{wall}}=\SI{10}{\centi\meter}$ & \SI{24}{\percent} & \SI{124}{\milli\second} & \SI{0.71}{\mega\ampere} \\
             $r_{\text{wall}}=\SI{30}{\centi\meter}$ & \SI{37}{\percent} & \SI{125}{\milli\second} & \SI{1.4}{\mega\ampere} \\
             $t_{\text{wall}}=\SI{500}{\milli\second}$ & \SI{29}{\percent} & \SI{210}{\milli\second} & \SI{3.9}{\mega\ampere} \\
             $t_{\text{wall}}=\SI{10}{\milli\second}$ & \SI{60}{\percent} & \SI{119}{\milli\second} & \SI{2.9}{\mega\ampere} \\
        \end{tabular}
        \caption{Current conversion, CQ time $t_{\text{CQ}}$, and remaining ohmic current at the end of the simulation (\SI{150}{\milli\second}) for cases differing from the baseline case in one input parameter, as listed in the first column.}
        \label{Tab:cc-tcq}
    \end{center}
\end{table}

\begin{figure}
    \begin{center}
    \includegraphics[width=\textwidth]{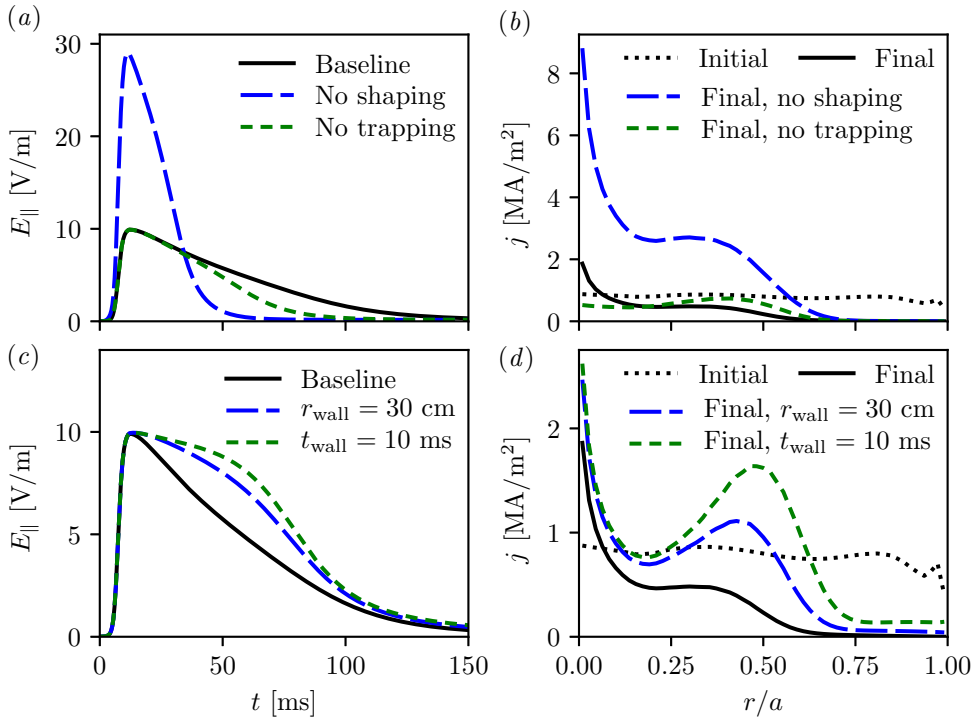}
    \end{center}
    \vspace{-\baselineskip}
    \caption{Electric field evolution at $r/a=0.5$ (\textit{a}, \textit{c}) and final current densities (\textit{b}, \textit{d}). Panels (\textit{a}, \textit{b}) compare the baseline (solid) with the \textit{no shaping} (long dashed) and \textit{no trapping} (short dashed) cases. Panels (\textit{c}, \textit{d}) compare the baseline ($r_{\text{wall}}=\SI{0}{\centi\meter}$, $t_{\text{wall}}=\infty$, solid) with the cases using $r_{\text{wall}}=\SI{30}{\centi\meter}$ (long dashed) and $t_{\text{wall}}=\SI{10}{\milli\second}$ (short dashed). In (\textit{b}, \textit{d}) the initial current density profile is also included (dotted).}
    \label{fig:E-j-variations}    
\end{figure}

The plasma shape will in reality evolve during a disruption, so we compare the runaway generation in the highly shaped baseline configuration to a case with no shaping, i.e.~a fixed circular cross-section with no Shafranov shift (the $j_0$ profile shown in figure~\ref{fig:input}\textit{a} is then multiplied by 3 to keep the total current constant), to gain an understanding of the impact that shaping can have on the disruption dynamics. The trapping corrections on the generation rates are excluded in one of the variations, as trapping can be expected to be strong in highly shaped compact tokamaks, making it interesting to see exactly how much it impacts the results. 
We note that both the case where trapping effects on the runaway generation mechanisms are excluded (\textit{no trapping}), as well as the case with a circular cross section (\textit{no shaping}), increase the current conversion and decrease the CQ time. The physical reasons behind these changes are different in the two cases. In figures \ref{fig:E-j-variations}\textit{a} and \ref{fig:E-j-variations}\textit{b} the electric field and the current density profiles of these two cases are shown, together with the baseline case for comparison. The maximum electric field is notably higher in the \textit{no shaping} case, which leads to the increased current conversion. In the \textit{no trapping} case the maximum electric field is the same but the current contributions at larger radii --- where a significant fraction of particles would otherwise be trapped --- are increased, as seen in the radial distribution of the final current density.

We have also investigated the effects of changing the wall distance $r_\text{wall}$ and the wall time $t_\text{wall}$. The magnetic fields in conducting structures have a complex response to a disruption, which depends on the geometry and material composition, and this response is currently determined by the two numbers in \textsc{dream}, $r_\text{wall}$ and $t_\text{wall}$. They can be estimated from measurements or detailed electromagnetic calculations, but due to the uncertainty in these values for a preliminary reactor design like BurST it is useful to let $r_\text{wall}$ and $t_\text{wall}$ vary, to understand the sensitivity of the result to the assumed values.  

\begin{figure}
    \begin{center}
    \includegraphics[width=\textwidth]{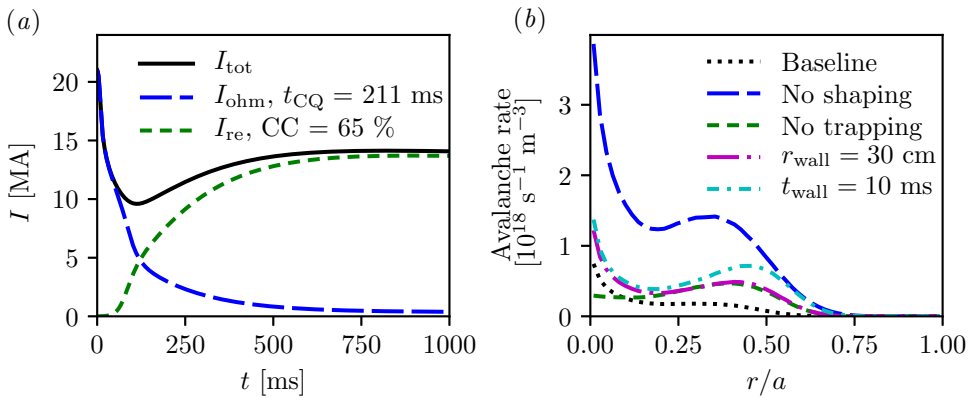}
    \end{center}
    \vspace{-\baselineskip}
    \caption{(\textit{a}) Total plasma current (solid) as a function of time, together with the ohmic (long dashed) and runaway (short dashed) contributions, when using a finite wall time $t_{\text{wall}}=\SI{500}{\milli\second}$. Note the long time scale plotted, and compare to figure~\ref{fig:1-15}\textit{a}. (\textit{b}) Characteristic radial profiles of the avalanche runaway rate: baseline case (dotted, see figure~\ref{fig:1-15}\textit{d}), \textit{no shaping} case (long dashed), \textit{no trapping} case (short dashed), $r_{\text{wall}}=\SI{30}{\centi\meter}$ case (long dash-dotted), and $t_{\text{wall}}=\SI{10}{\milli\second}$ case (short dash-dotted). }
    \label{fig:finite-walltI-ava-cases}
\end{figure}

By increasing the wall distance from zero to $10\,\rm cm$ then to $30\,\rm cm$, 
or reducing the wall time from that of a perfectly conducting wall ($t_{\text{wall}}=\infty$) to an ITER-like value $t_{\text{wall}}=0.5\,\rm s$ and an even shorter value of $10\,\rm ms$, we find an increased current conversion, as well as increased $t_{\text{CQ}}$. The electric field and the current density for these cases can be seen in figures \ref{fig:E-j-variations}\textit{c} and \ref{fig:E-j-variations}\textit{d}. In both cases the off-axis contributions are increased compared to the baseline; in the  $r_{\text{wall}}>0$ case this is due to magnetic energy returning to the plasma from the vacuum region between the plasma and the wall, while in the $t_{\text{wall}}<\infty$ case it is magnetic energy from the wall and potentially the surrounding structures that diffuses back into the plasma. The stronger electric fields near the edge affect the generation mechanisms, such that more runaway production happens further away from the centre of the plasma. This is visible in the final current densities, which have a larger off-axis contribution. 
Another interesting observation concerning changing to a finite wall time is that the dynamics of the total current changes as a result of magnetic energy returning to the plasma from the wall. The effect of this is that a runaway plateau phase is not immediately reached after the CQ, but there is rather a long gradual increase in the runaway current before the plateau is reached, see figure~\ref{fig:finite-walltI-ava-cases}a. The time needed to reach the plateau in this case is much longer than $\SI{150}{\milli\second}$, which is the time-scale where the control over the plasma is deemed lost after a disruption. This implies that the maximum runaway current would depend strongly on when the control over the plasma is lost, as compared to the case of a perfectly conducting wall.

The final column of table~\ref{Tab:cc-tcq} indicates the remaining ohmic current at the end of the simulations. Values of more than tens of kiloamperes indicate an incomplete current quench and we see that this is the situation in all variations evaluated. As noted above, the remaining ohmic current could become a problem, because it might be converted to runaway current if control over the plasma is lost at some later time. The case with the most residual ohmic current is that with a wall time of \SI{500}{\milli\second} (almost 9 times that in the baseline) and the case with the least is the \textit{no shaping} case (around 6 times lower than that in the baseline). This indicates that the strong ST shaping makes the residual ohmic current problem worse, whilst it reduces the initial runaway problem.

Regarding the different generation mechanisms, we have already touched upon the weak influence of the tritium seed, as confirmed by the cases where the amount of tritium is varied. In all other cases tritium generation remains of the same order of magnitude. When and where the tritium runaway generation happens does however change, an example being the radial gap at around $r/a \approx 0.25$ in figure~\ref{fig:1-15-mech}\textit{b} disappearing in the \textit{no shaping} and \textit{no trapping} cases. 
In all cases in table~\ref{Tab:cc-tcq} hot-tail is the dominant primary generation mechanism, whereas most of the runaway current is generated through the avalanche. As the parameters are varied, the only case which produces a change in the hot-tail generation compared to figure~\ref{fig:1-15-mech}\textit{c} is the \textit{no shaping} case, where the electric field is stronger. In this case the hot-tail generation happens at the same radii during the same time period, but the maximum value is increased by a factor of around five. As the avalanche multiplication dominates in all studied cases, it determines the shape of the final current density, which can be seen by comparing the profile of the avalanche rates for the parameter variations studied, as shown in figure~\ref{fig:finite-walltI-ava-cases}\textit{b}, to the final current profiles in figures~\ref{fig:E-j-variations}\textit{b} and \ref{fig:E-j-variations}\textit{d}.

Finally, we considered the unmitigated evolution for BurST profiles which were not optimised for energy confinement (dotted lines in figures~\ref{fig:input}\textit{a}-\textit{c}). With other parameters taken to be the same as the baseline case, the current conversion increased to \SI{22}{\percent}, compared to \SI{14}{\percent} in the baseline case. This means that the profiles optimised for energy confinement also reduce the runaway production. We repeated the study of the effect of the parameter variations listed in table~\ref{Tab:cc-tcq} for these un-optimised profiles and found the same trends in the runaway current conversion; the runaway current conversion stays the same or increases. This indicates that there is a robustness in the obtained results with respect to changes of this type in the input profiles.

\section{Mitigation with Massive Material Injection}
\label{sec:res_mitigated}
In the previous section we found that reactor-scale ST disruptions can be expected to generate significant runaway populations, not atypical of disruptions in conventional reactor scale tokamaks \citep{iter_matinj}. Therefore, in this section we undertake a first study of the effectiveness of straightforward material injection mitigation on the runaway dynamics. The considered mitigation strategy is injection of a large quantity of mixed deuterium and neon. 
There are different schemes for injecting material into the plasma during a disruption. {In the massive gas injection (MGI) scheme neutral gas is released into the plasma from a vault in the tokamak wall \citep{Hollmann}. Despite its relative simplicity, this method has a disadvantage: the material begins to ionise at the edge of the plasma as soon as it is injected, thus becoming magnetically confined before reaching and cooling the hottest central parts of the plasma. This issue is overcome in the shattered pellet injection (SPI) scheme --- the baseline disruption mitigation technique on ITER \citep{LehnenIAEA} --- where frozen pellet shards are injected into the plasma.} Here, the details of the material delivery are not considered and we work from assumed deposited material profiles. This also facilitates comparison of the results to material injection disruption mitigation effects studied previously in conventionally shaped tokamaks.

For the material injection simulations the magnetic perturbation $\delta B/B$ is estimated from the values for $t_0$ and $T_f$ during the TQ, such that the decay time scale before the radiative collapse is approximately $t_0$. The transport is active during the time it would take for the temperature to decay exponentially from the initial temperature to \SI{100}{\electronvolt}, according to equation~(\ref{eq:expT}), after which the MHD-induced losses represented by this perturbation generally no longer dominate. The estimate for the time over which this is active is conservative, as dilution, ionisation and radiation losses generally make the total thermal quench time shorter than this. Also, in reality, the flux surfaces tend to re-heal after the plasma has lost most of its thermal energy \citep{Sommariva_2018}, so the electron heat diffusivity would drop rapidly. Our conservative estimate ensures that the transport is active here during the entire TQ. Having significant heat diffusivity after the plasma has reached a low quasi-equilibrium temperature has in fact very little effect, as then radiative heat losses dominate by a large factor.

Trapping corrections to the growth rates are turned off for the material injection simulations, as these effects are negligible at high densities and in the presence of significant impurity content, due to the high collisionality. Also note that trapping is less important in the presence of very high electric fields, as particles can be accelerated out from the trapped region faster than their orbit time \citep{McDevitt_2019}. The same assumption was made by \citet{iter_matinj} when modelling mitigation in ITER-like plasmas.

We identify trends in the impact of material mitigation by scanning the injected deuterium and neon densities over the ranges 
$n_{\text{D}}=10^{20}\,\si{\meter^{-3}}- 10^{22}\,\si{\meter^{-3}}$ and $n_{\text{Ne}}=10^{16}\,\si{\meter^{-3}}- 10^{20}\,\si{\meter^{-3}}$, which are similar to those previously used by \citet{iter_matinj}.
\begin{figure}
    \begin{center}
    \includegraphics[width=\textwidth]{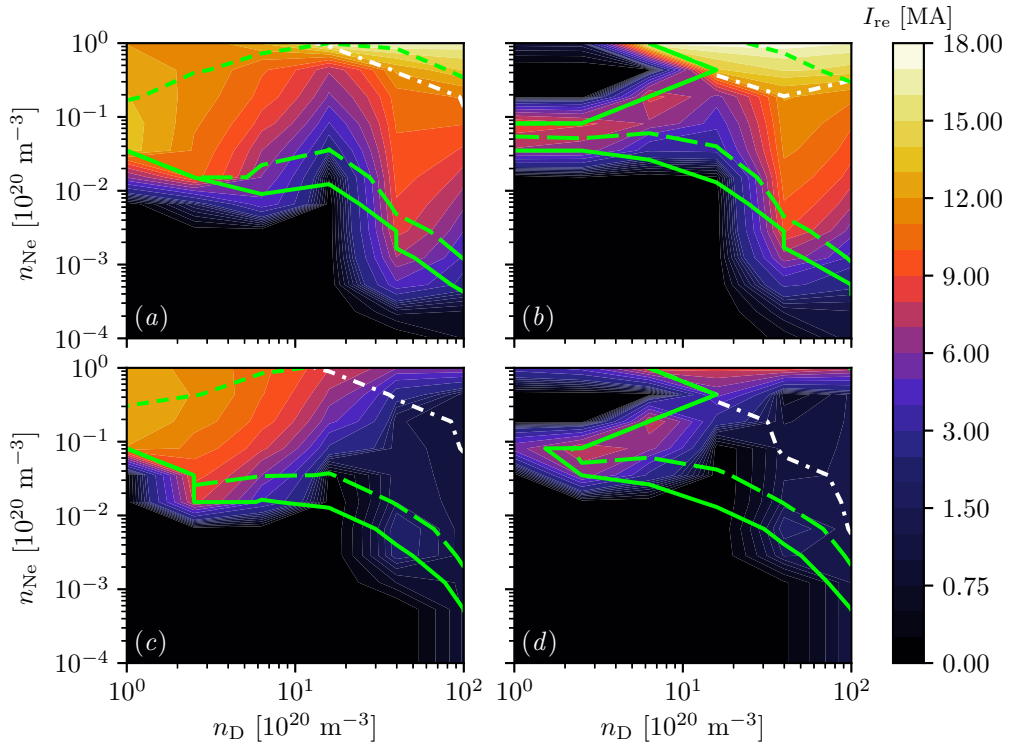}
    \end{center}
    \vspace{-\baselineskip}
    \caption{Maximum runaway current $I_{\text{re}}$ as a function of injected deuterium ($n_{\text{D}}$) and neon ($n_{\text{Ne}}$) densities, in the case where (\textit{a}, \textit{c}) $\delta {B}/B\approx\SI{0.6}{\percent}$ and (\textit{b}, \textit{d}) $\delta {B}/B\approx\SI{0.2}{\percent}$. Panels (\textit{a}, \textit{b}) include only heat transport whereas panels (\textit{c}, \textit{d}) include runaway transport as well. Between the short- and long-dashed lines, $t_{\text{CQ}}$ takes values between \SI{20}{\milli\second} and \SI{100}{\milli\second}. To the left of the solid line, $t_{\text{CQ}}$ is longer than \SI{150}{\milli\second} or the CQ is incomplete (in which case the CQ time would be much longer than \SI{150}{\milli\second}). Above the dash-dotted line the transported fraction of the thermal energy loss is lower than \SI{10}{\percent}.}  
    \label{fig:mit_scan_highdBB}
\end{figure} 
We consider mitigation of two cases, a high transport case with $\delta {B}/B\approx\SI{0.6}{\percent}$ representative of a fast TQ ($t_0=\SI{1}{\milli\second}$, $ T_f=\SI{10}{\electronvolt}$) and a low transport case {with $\delta {B}/B\approx\SI{0.2}{\percent}$} representative of a slow TQ ($t_0=\SI{10}{\milli\second}$, $ T_f=\SI{10}{\electronvolt}$). The magnetic perturbations determining the transport were estimated based on the temperature decay parameters. With $t_0=\SI{1}{\milli\second}$, $ T_f=\SI{10}{\electronvolt}$ the corresponding unmitigated case has a current conversion of \SI{46}{\percent}, or equivalently a maximum runaway current of \SI{9.7}{\mega\ampere}, and $t_{\text{CQ}} = \SI{75}{\milli\second}$. Therefore, the mitigation objective in this case is to reduce the runaway current while keeping the CQ time within the limits. With $t_0=\SI{10}{\milli\second}$, $ T_f=\SI{10}{\electronvolt}$ the corresponding unmitigated case has a vanishingly small current conversion, but an incomplete CQ with an ohmic current of \SI{8.7}{\mega\ampere} remaining at the end of the simulation. The goal of mitigation in this case is thus instead to make the ohmic current decay faster to obtain $t_{\text{CQ}}$ between the limits, while keeping the runaway current at a low level.

The resulting maximum runaway current as a function of injected deuterium and neon is shown in figures~\ref{fig:mit_scan_highdBB}\textit{a} and~\ref{fig:mit_scan_highdBB}\textit{c}  for the case with high transport and in figures~\ref{fig:mit_scan_highdBB}\textit{b} and~\ref{fig:mit_scan_highdBB}\textit{d} for the case with low transport.
In figures~\ref{fig:mit_scan_highdBB}\textit{a} and~\ref{fig:mit_scan_highdBB}\textit{b} we show the result only accounting for diffusive heat transport due to the magnetic perturbations, whilst figures~\ref{fig:mit_scan_highdBB}\textit{c} and~\ref{fig:mit_scan_highdBB}\textit{d} also include the consistent diffusive radial transport of runaways --- this can be seen to have a significant impact on the operating space.  In these simulations the magnetic perturbations were applied only during the {time} it took for the temperature in the equivalent cases in figures~\ref{fig:mit_scan_highdBB}\textit{a} and ~\ref{fig:mit_scan_highdBB}\textit{b} to fall to 100eV, and are therefore less conservative. 
The figures include several boundaries for reference: lower $t_{\text{CQ}}$ boundary of \SI{20}{\milli\second} (short dashed, producing no constraint in figure~\ref{fig:mit_scan_highdBB}\textit{d}), two possible upper $t_{\text{CQ}}$ boundaries of \SI{100}{\milli\second} (long dashed) and \SI{150}{\milli\second} (solid), as well as a boundary indicating \SI{10}{\percent} of the thermal energy being lost through radial transport (short dash-dotted). As can be seen in all cases, there is a region of low runaway current (below \SI{0.5}{MA}) for low injected densities, as well as at high neon and low deuterium density injected in the low transport case. In these regions of low runaway current the ohmic CQ is incomplete. This is evident from the regions being below or to the left of the solid lines, indicating that the injected deuterium-neon mixture is insufficient to induce a complete radiative collapse.

In figures~\ref{fig:mit_scan_highdBB}\textit{a} and~\ref{fig:mit_scan_highdBB}\textit{b}, i.e. neglecting particle loss due to the magnetic perturbations, we see that in the region between the lower $t_{\text{CQ}}$ limit of \SI{20}{\milli\second} and the upper $t_{\text{CQ}}$ limit of \SI{100}{\milli\second} runaway currents between \SI{2.5}{\mega\ampere} and \SI{18}{\mega\ampere} are obtained. If the upper CQ time limit can be extended to \SI{150}{\milli\second}, the region contains runaway currents down to \SI{0}{\mega\ampere}, around injected densities of $n_{\text{D}}\approx1.6\cdot10^{21}\,$\si{\meter^{-3}} and $n_{\text{Ne}}\approx1.5\cdot10^{18}\,$\si{\meter^{-3}}.
However, we caution that the largest increase in the avalanche multiplication is here observed at the very end of the simulation. This is in accordance with \citep{Hesslow_fluid_ava}, where the runaway generation is first suppressed by material injection, to later become large due to a stronger avalanche in the presence of heavy impurities. There is also about \SI{1.4}{\mega\ampere} ohmic current left, which thus could readily be converted to a runaway current if the plasma survived beyond the \SI{150}{\milli\second} mark.

We look in more detail at the region in figure~\ref{fig:mit_scan_highdBB}\textit{a} where the best performance in terms of low runaway current is being obtained, that is $n_{\text{D}}\approx1.6\cdot10^{21}\,$\si{\meter^{-3}} and $n_{\text{Ne}}\approx1.5\cdot10^{18}\,$\si{\meter^{-3}}, and find that the tritium, hot-tail and avalanche generation all behave differently than in the corresponding unmitigated case. Hot-tail is still the dominant primary seed, although the maximum generation rate is reduced from $\sim10^{16}\,\si{\second^{-1}\meter^{-3}}$ to  $\sim10^{7}\,\si{\second^{-1}\meter^{-3}}$. Also, the generation only occurs over about \SI{0.5}{\milli\second}, compared to \SI{2}{\milli\second} in the unmitigated case, and is localised inside a more limited radial range around the plasma centre (up to $r/a\approx0.6$ rather than $r/a\approx1.0$). 
With this combination of injected deuterium-neon densities the fast electrons that formed the hot-tail in the unmitigated case slow down more before the electric field rises, leading to a smaller ``tail'' that can be converted to runaways. The tritium seed is fully suppressed, which indicates that the critical runaway energy for generation through tritium decay is increased for this level of impurity injection, in accordance with results by \citet{iter_matinj}. The smaller runaway seed currents lower the outcome of the avalanche multiplication, leading to the observed low runaway current.

When we now include particle loss due to the magnetic perturbations, figures~\ref{fig:mit_scan_highdBB}\textit{c} and~\ref{fig:mit_scan_highdBB}\textit{d}, a significant region opens up with both tolerable ohmic current evolution and runaway current.
In the case of the fast TQ, where mitigation is needed to reduce the runaway current while keeping the CQ time within acceptable limits (figure~\ref{fig:mit_scan_highdBB}\textit{c}), this region occurs for $n_{\text{D}}$ above around $4\cdot10^{21}\,$\si{\meter^{-3}} and $n_{\text{Ne}}$ in the range $1 - 20 \cdot10^{18}\,$\si{\meter^{-3}}. In the case of a slow TQ, where mitigation is needed to reduce the CQ time while keeping the runaway current low (figure~\ref{fig:mit_scan_highdBB}\textit{d}), this region appears for $n_{\text{D}} \approx 1.5 -3.5\cdot10^{21}\,$\si{\meter^{-3}} with $n_{\text{Ne}} \approx 2 - 15 \cdot10^{18}\,$\si{\meter^{-3}}, then widens for $n_{\text{D}}$ above $3.5\cdot10^{21}\,$\si{\meter^{-3}} to include $n_{\text{Ne}} \approx 1 - 40 \cdot10^{18}\,$\si{\meter^{-3}}. In these regions, the runaway currents range from \SI{0}{\mega\ampere} to about \SI{1.5}{\mega\ampere}.
Looking again in detail at the point in figure~\ref{fig:mit_scan_highdBB}\textit{c} with  $n_{\text{D}}\approx1.6\cdot10^{21}\,$\si{\meter^{-3}} and $n_{\text{Ne}}\approx1.5\cdot10^{18}\,$\si{\meter^{-3}}, we find that inclusion of the particle transport has reduced the maximum hot-tail generation rate to $\sim10^{5}\,\si{\second^{-1}\meter^{-3}}$. This results in a weaker avalanche and the low runaway currents in this region.
This process is responsible for opening the whole region of acceptable evolution in figures~\ref{fig:mit_scan_highdBB}\textit{c} and~\ref{fig:mit_scan_highdBB}\textit{d}.

Finally, we take the limit on transported thermal energy loss into account, as well as the $t_{\text{CQ}}$ limits. We see from figures~\ref{fig:mit_scan_highdBB}\textit{a} and~\ref{fig:mit_scan_highdBB}\textit{b} that without accounting for particle transport by the magnetic perturbations there is no region where it is possible to fulfil all three demands simultaneously. In the regions above the short dash-dotted line in these figures, where the transported fraction is below \SI{10}{\percent}, the runaway current is at least \SI{11}{\mega\ampere}. 
Including the particle transport only moderately affects the energy loss boundary in the slow TQ case, but the reduced runaway current generation does offer acceptable parameter spaces at high injected levels of deuterium and neon. 
The reason behind the transported fraction reaching high levels seems to be the shape of the initial temperature and density profiles, see figures~\ref{fig:input}\textit{b}-\textit{c}. For example, in the case with the lower $\delta B/B$ for $n_{\text{D}}\approx1.6\cdot10^{21}\,$\si{\meter^{-3}} and $n_{\text{Ne}}\approx8.1\cdot10^{18}\,$\si{\meter^{-3}}, the transported fraction is \SI{61}{\percent}. If we instead change to a flat initial density of $n_{e0} =10^{20}\,\si{\meter^{-3}}$,  as in \citep{iter_matinj}, the transported heat loss fraction is reduced to about \SI{33}{\percent}. If also the temperature profile is changed to that of \citep{iter_matinj}, i.e. $T_0(r) = 20\cdot[1-(r/a)^2]\,\si{\kilo\electronvolt}$, the fraction is further reduced to \SI{14}{\percent}, very close to the acceptable level.

\section{Discussion and Conclusions}
\label{sec:conclusions}
We have shown that mitigation would typically be required to {keep the runaway} current below acceptable levels in reactor-scale STs, and that the runaway generation is comparable to that seen in conventional tokamak reactor scenarios. 
The dominant primary generation mechanism is hot tail, that reaches 6 (18) orders of magnitude higher values than the tritium decay (Dreicer) seed, and the avalanche gain is very high, as expected for a high current tokamak.   

Our simulations show that both removing the shaping, as well as removing the trapping corrections, increase the runaway current. When removing the shaping the increase was due to a higher electric field, while the increase when removing the trapping correction was due to more current contributing at larger radii. {This may naively suggest that if the plasma evolves from its highly shaped initial state through to a smaller, more circular configuration as the outer layers of plasma are possibly lost during the disruption, the runaway generation would increase compared to the results given here at constant shape. It is likely more complicated though, as then the current density would not increase in the plasma core in such a process, while the scraped off current may be re-induced at the boundary of the plasma with flux surfaces.}  Increasing the wall distance or decreasing the wall time also lead to an increased runaway current. In both cases this increase was due to an increased energy reservoir, either from the vacuum between the plasma and the wall, or from the surrounding structures.

Whilst successful control of the ohmic current evolution and runaway currents seems to be within reach in the case of disruption mitigation by mixed deuterium-neon injection, quantities at the higher end of the range explored are required to control the transported energy level, due to the peaked density and large temperature pedestal.

There are several aspects of our analysis that could be expanded on to move towards quantitative predictions of the runaway generation in ST-based fusion reactors. Importantly, the effect of Compton seed due to vessel activation has previously been found to have a significant effect on runaway generation in conventional tokamaks due to Compton scattering. This was demonstrated for example by \citet{iter_matinj}, where adequate mitigation for ITER-like parameters could not be achieved with mixed {material injection} in the activated phase of operations. Such modelling requires the spectrum of $\gamma$-photons emitted from the plasma-facing components. Given for ITER in \citep{MartinSolis2017}, this remains to be determined for alternative reactor designs. However, the runaway current in high-current devices is only a logarithmically weak function of the seed \citep{iter_matinj}, so it might not have a major effect on the result, although it could have a more significant impact if the hot-tail and Dreicer seeds would be lost through transport during the TQ.

We have studied here the two limiting cases where either the same shaping was kept throughout the disruption or a circular cross-section was assumed throughout. 
{\textsc{dream}} does currently not allow for the time evolution of shaping parameters, but it may be possible to implement by enabling the user to prescribe time-dependent shaping parameters.

Regarding the material injection it could be interesting to study another impurity, such as argon. Using a heavier impurity might lead to more radiation and thus possibly reduce the high transported fraction of the thermal energy loss, but at the same time we expect that it has the potential to lead to a larger avalanche, due to the larger number of bound target electrons \citep{Hesslow_fluid_ava}. Also, relaxing the assumption of a flat density profile for the injected material would improve the model, as in reality the material is injected at the wall and in most cases begins to ionise immediately \citep{Svenningsson_PRL}. Additionally, SPI (shattered pellet injection) might be considered instead of the neutral gas injection modelled here (uniformly distributed at $t=0$), as SPI has the potential to reach and cool the centre of the plasma \citep{Vallhagen_SPI}.

\section*{Acknowledgements} 
The authors are grateful to T Hender, B Patel, H Bergström, P Halldestam, L-G Eriksson and E Nardon for fruitful discussions. This work was supported by the Swedish Research Council (Dnr.~2018-03911) and part-funded by the EPSRC Energy Programme [grant number EP/W006839/1] and the Swiss National Science Foundation. The work has been carried out within the framework of the EUROfusion Consortium, funded by the European Union via the Euratom Research and Training Programme (Grant Agreement No 101052200 — EUROfusion). Views and opinions expressed are however those of the author(s) only and do not necessarily reflect those of the European Union or the European Commission. Neither the European Union nor the European Commission can be held responsible for them.

\appendix

\section{Details of the Input Parameters}
\label{sec:det}
\begin{figure}
    \begin{center}
    \includegraphics[width=\textwidth]{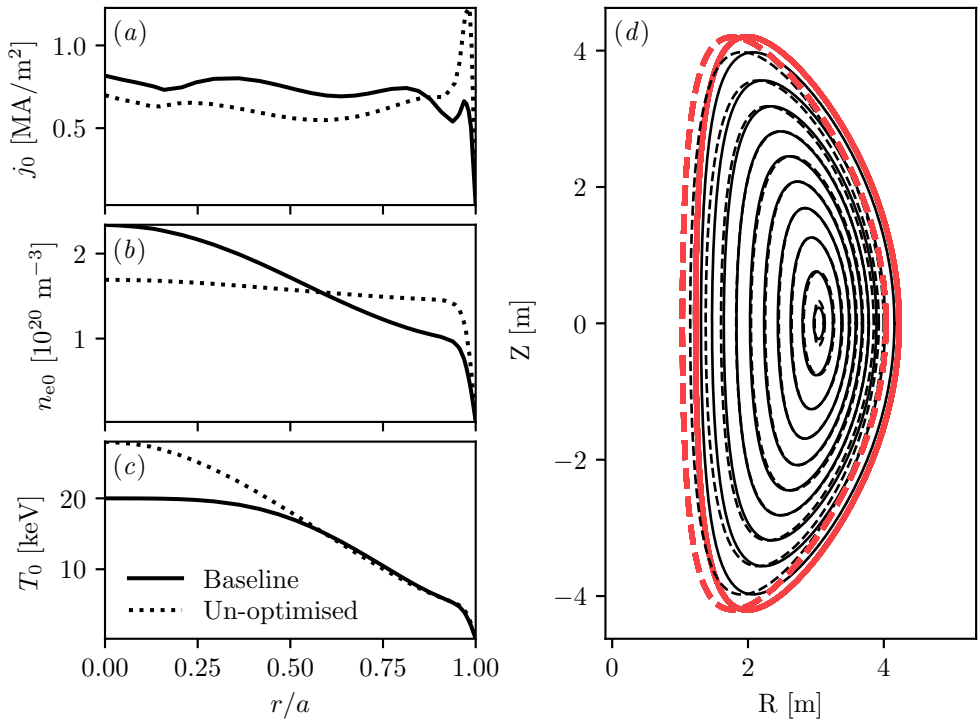}
    \end{center}
    \vspace{-\baselineskip}
    \caption{Radial profiles for initial (\textit{a}) current density, (\textit{b}) electron density and (\textit{c}) temperature, both for the baseline case and the profiles not optimised for energy confinement.  (\textit{d}) Shape of the input equilibrium flux surfaces. The dashed lines correspond to the unmodified equilibrium and the solid lines indicate the modified flux surfaces used in the simulations. The thicker red lines mark the outermost flux surface in each case.
    } 
    \label{fig:input}
\end{figure}

The BurST profiles for the initial total current density, electron density and temperature used in our simulations are shown in figure~\ref{fig:input}\textit{a}-\textit{c} \citep{Bhavin_phd}. The total initial plasma current is $I_{\text{tot},0}=\SI{21}{\mega\ampere}$. 
Solid lines show the baseline profiles, which were obtained by optimising the energy confinement with respect to microinstabilities in BurST. An example of un-optimised profiles, shown as dotted lines in figure~\ref{fig:input}\textit{a}-\textit{c}, are used here to study the impact of changes in the input profiles on the results. 
In figure~\ref{fig:input}\textit{d} the flux surfaces of the input equilibrium are shown. The plasma major and minor radii are $R_0=\SI{3.05}{\meter}$ and $a=\SI{1.5}{\meter}$, respectively, and the elongation at the outermost flux surface is $\kappa(a)=2.8$.  

In a spherical tokamak the poloidal and toroidal magnetic field components can be comparable, so the magnetic field strength may have more than one minimum on a flux surface \citep{Wilson_2004}, which is the case here. As this is incompatible with the requirements on the input by \textsc{dream}, in our simulations the shaping profiles have been slightly modified to avoid multiple magnetic field minima. The double magnetic field strength minima on the outer flux surfaces is caused by the large Shafranov shift,  $\Delta$, which is about \SI{-0.5}{\meter} at the edge. We empirically found that limiting the edge value of the Shafranov shift to \SI{-0.3}{\meter} is sufficient to avoid the issue of double minima. Therefore we modified the slope of the Shafranov shift profile for large $r$ to obtain $\Delta(r=a)\approx\SI{-0.3}{\meter}$.  As we keep $R_0$ fixed, this implies a shift of the outermost flux surface, since $\Delta(r=0)=0$ according to the definition of the Shafranov shift in \textsc{dream}. The flux surfaces of the unmodified equilibrium are shown as dashed lines. By running simulations for varying radial location of the outermost flux surface, we confirmed that the sensitivity of the results to the shift is negligible. 

\bibliographystyle{jpp}
\bibliography{bibliography}

\end{document}